\documentclass[12pt]{cernart}
\usepackage{amssymb}
\usepackage{amsmath}
\usepackage{graphicx}
\usepackage{multicol}

\addtocounter{footnote}{1}

\begin{document}

\begin{titlepage}
\docnum{CERN--PH--EP/2005--049}
\hbox to \hsize{\hskip123mm\hbox{27 October 2005}\hss}

\vspace{1cm}
\title{\Large Gluon polarization in the nucleon from quasi-real photoproduction of  high-$p_{\rm T}$ hadron pairs}

\begin{center}{\large  The COMPASS Collaboration}\\
\end{center}

\vspace{3cm}
\begin{abstract}
We present a determination of the gluon polarization $\Delta G/G$
in the nucleon, based on the helicity asymmetry of quasi-real photoproduction
events, $Q^2<1\ ({\rm GeV}/c)^2$, with a pair of large transverse-momentum hadrons in the final state. The data
were obtained by the COMPASS experiment at CERN using a 160 GeV
polarized muon beam scattered on a polarized $^6$LiD target.
The helicity asymmetry for the selected
events is 
$\langle A_{\parallel}/D \rangle = 0.002 \pm 0.019({\rm stat.}) \pm 0.003({\rm syst.})$.
From this value, we obtain in a leading-order QCD analysis
$\Delta G/G=0.024 \pm 0.089({\rm stat.})\pm 0.057({\rm syst.})$ at
$x_g = 0.095$ and $\mu^2 \simeq 3 \ ({\rm GeV}/c)^2$.
\vfill
\submitted{(Submitted to Physics Letters B)}
\end{abstract}

\dimen\footins=9in
\begin{Authlist}
E.S.~Ageev\Iref{protvino},
V.Yu.~Alexakhin\Iref{dubna},
Yu.~Alexandrov\Iref{moscowlpi},
G.D.~Alexeev\Iref{dubna},
A.~Amoroso\Iref{turin},
B.~Badelek\Iref{warsaw},
F.~Balestra\Iref{turin},
J.~Ball\Iref{saclay},
G.~Baum\Iref{bielefeld},
Y.~Bedfer\Iref{saclay}
% ...
\end{Authlist}

\Instfoot{bielefeld}{ Universit\"at Bielefeld, Fakult\"at f\"ur Physik, 33501 Bielefeld, Germany\Aref{d}}
\Instfoot{bochum}{ Universit\"at Bochum, Institut f\"ur Experimentalphysik, 44780 Bochum, Germany\Aref{d}}
\Instfoot{bonniskp}{ Universit\"at Bonn, Helmholtz-Institut f\"ur  Strahlen- und Kernphysik, 53115 Bonn, Germany\Aref{d}}
\Instfoot{bonnpi}{ Universit\"at Bonn, Physikalisches Institut, 53115 Bonn, Germany\Aref{d}}
\Instfoot{brno}{Institute of Scientific Instruments, AS CR, 61264 Brno, Czech Republic\Aref{e}}
\Instfoot{burdwan}{ Burdwan University, Burdwan 713104, India\Aref{g}}
\Instfoot{calcutta}{ Matrivani Institute of Experimental Research \& Education, Calcutta-700 030, India\Aref{h}}
\Instfoot{dubna}{ Joint Institute for Nuclear Research, 141980 Dubna, Moscow region, Russia}
\Instfoot{erlangen}{ Universit\"at Erlangen--N\"urnberg, Physikalisches Institut, 91054 Erlangen, Germany\Aref{d}}
\Instfoot{freiburg}{ Universit\"at Freiburg, Physikalisches Institut, 79104 Freiburg, Germany\Aref{d}}
\Instfoot{cern}{ CERN, 1211 Geneva 23, Switzerland}
\Instfoot{heidelberg}{ Universit\"at Heidelberg, Physikalisches Institut,  69120 Heidelberg, Germany\Aref{d}}
\Instfoot{helsinki}{ Helsinki University of Technology, Low Temperature
                     Laboratory, 02015 HUT, Finland  and University of Helsinki, Helsinki
                     Institute of  Physics, 00014 Helsinki, Finland}
\Instfoot{licerec}{Technical University in Liberec, 46117 Liberec, Czech Republic\Aref{e}}
\Instfoot{lisbon}{ LIP, 1000-149 Lisbon, Portugal\Aref{f}}
\Instfoot{mainz}{ Universit\"at Mainz, Institut f\"ur Kernphysik, 55099 Mainz, Germany\Aref{d}}
\Instfoot{miyazaki}{University of Miyazaki, Miyazaki 889-2192, Japan\Aref{i}}
\Instfoot{moscowlpi}{Lebedev Physical Institute, 119991 Moscow, Russia}
\Instfoot{munichlmu}{Ludwig-Maximilians-Universit\"at M\"unchen, Department f\"ur Physik, 80799 Munich, Germany\Aref{d}}
\Instfoot{munichtu}{Technische Universit\"at M\"unchen, Physik Department, 85748 Garching, Germany\Aref{d}}
\Instfoot{nagoya}{Nagoya University, 464 Nagoya, Japan\Aref{i}}
\Instfoot{praguecu}{Charles University, Faculty of Mathematics and Physics, 18000 Prague, Czech Republic\Aref{e}}
\Instfoot{praguectu}{Czech Technical University in Prague, 16636 Prague, Czech Republic\Aref{e}}
\Instfoot{protvino}{ State Research Center of the Russian Federation, Institute for High Energy Physics, 142281 Protvino, Russia}
\Instfoot{saclay}{ CEA DAPNIA/SPhN Saclay, 91191 Gif-sur-Yvette, France}
\Instfoot{telaviv}{ Tel Aviv University, School of Physics and Astronomy, 
              69978 Tel Aviv, Israel\Aref{j}}
\Instfoot{triestictp}{ ICTP--INFN MLab Laboratory, 34014 Trieste, Italy}
\Instfoot{triest}{ INFN Trieste and University of Trieste, Department of Physics, 34127 Trieste, Italy}
\Instfoot{turin}{ INFN Turin and University of Turin, Physics Department, 10125 Turin, Italy}
\Instfoot{warsaw}{ Soltan Institute for Nuclear Studies and Warsaw University, 00-681 Warsaw, Poland\Aref{k} }
\Instfoot{warsawtu}{ Warsaw University of Technology, Institute of Radioelectronics, 00-665 Warsaw, Poland\Aref{l} }
\Anotfoot{a}{Also at IST, Universidade T\'ecnica de Lisboa, Lisbon, Portugal}
\Anotfoot{b}{Also at University of East Piedmont, 15100 Alessandria, Italy}
\Anotfoot{c}{On leave of absence from JINR Dubna}               
\Anotfoot{d}{Supported by the German Bundesministerium f\"ur Bildung und Forschung}
\Anotfoot{e}{Suppported by Czech Republic MEYS grants ME492 and LA242}
\Anotfoot{f}{Supported by the Portuguese FCT - Funda\c{c}\~ao para
               a Ci\^encia e Tecnologia grants POCTI/FNU/49501/2002 and POCTI/FNU/50192/2003}
\Anotfoot{g}{Supported by UGC-DSA II grants, Govt. of India}
\Anotfoot{h}{Supported by  the Shailabala Biswas Education Trust}
\Anotfoot{i}{Supported by the Ministry of Education, Culture, Sports,
               Science and Technology, Japan}
\Anotfoot{j}{Supported by the Israel Science Foundation, funded by the Israel Academy of Sciences and Humanities}
\Anotfoot{k}{Supported by KBN grant nr 621/E-78/SPUB-M/CERN/P-03/DZ 298 2000 and
               nr 621/E-78/SPB/CERN/P-03/DWM 576/2003-2006, and the MNII research funds for 2005-2007}
\Anotfoot{l}{Supported by  KBN grant nr 134/E-365/SPUB-M/CERN/P-03/DZ299/2000}
\Anotfoot{m}{Also at Gesellschaft f\"ur Schwerionenforschung, Darmstadt, Germany}
\vfill

\hbox to 0pt {~}
\end{titlepage}
\setcounter{footnote}{0}
\section{Introduction}

The decomposition of the nucleon spin in terms of the contributions from
its constituents has been a central topic in
polarized deep-inelastic scattering (DIS) for the last twenty years. The
European Muon Collaboration study of the proton spin
structure~\cite{EMC:1988} has shown that the spin of the quarks only
contributes to a small fraction $\Delta \Sigma$ of the proton spin. This
result has been
confirmed by several experiments on the proton, the deuteron, and
$^3$He, establishing $\Delta \Sigma$ between 20\% and
30\%~\cite{Adeva:1998vv,Anthony:2000fn}, in contrast to the 60\%
expected in the quark-parton model~\cite{Ellis:1973kp}. 

Another contribution to the nucleon spin, $\Delta G$, originates from
the spin of the gluons. In inclusive DIS, it can only be determined from
the $Q^2$ dependence of the spin structure function $g_1$. 
Next-to-Leading Order (NLO) QCD analyses provide estimates for 
$\Delta G$ below or around unity at a scale of 3~$({\rm GeV}/c)^2$. The
precision of these fits is however strongly limited by the small $Q^2$
range covered by the data.

In semi-inclusive DIS or in proton--proton scattering, the final state
can be used to select hard processes involving gluons from the nucleon.
In polarized semi-inclusive DIS, the polarization $\Delta G/G$ of gluons
carrying a fraction $x_{\rm g}$ of the nucleon momentum is obtained from
the cross-section helicity asymmetry of the photon--gluon fusion (PGF),
$\gamma^* g \rightarrow q \bar q$. 

Two procedures have been proposed to tag this process. The first one
consists in selecting open-charm events, which provides the purest
sample of PGF events~\cite{Watson:1982,Gluck:1988uj}, but at a low rate.
Another possibility is to select events with two jets at high transverse
momentum, $p_{\rm T}$, with respect to the virtual photon direction or,
in fixed-target experiments, two high-$p_{\rm T}$
hadrons~\cite{Bravar:1998kb}.
The latter procedure provides much larger statistics but leaves a
significant fraction of background events in the selected sample. As a
result, the cross-section helicity asymmetry $A_\parallel$ contains in
addition to the contribution from PGF a contribution $A_{\rm bgd}$ from
the background processes:
\begin{equation}
  A_\parallel  = R_{\rm PGF}\ \hat a_{\rm LL}^{\rm PGF}  \frac{\Delta G}{G} +  A_{\rm bgd}.
  \label{eq:aparod_decomppgf}
\end{equation}
Here, $R_{\rm PGF}$ is the fraction of PGF events and $\hat a_{\rm
LL}^{\rm PGF} \equiv {\rm d}\Delta \sigma_{\rm PGF}^{\mu g} / {\rm
d}\sigma_{\rm PGF}^{\mu g}$ is the  analyzing power of PGF that is the
helicity asymmetry of the hard lepton--gluon scattering cross-section.
This quantity is calculated from the leading order expressions of the
polarized and unpolarized partonic cross-sections. On the other hand,
$R_{\rm PGF}$ and $A_{\rm bgd}$ must be estimated by a simulation, which
introduces a model dependence in the evaluation of $\Delta G/G$.

This paper presents a measurement of the cross-section helicity
asymmetry obtained for the large sample of muon--deuteron events
collected by the COMPASS experiment at CERN in the low virtuality
domain, $Q^2<1$~$({\rm GeV}/c)^2$. We select interactions in which a
pair of high-$p_{\rm T}$ hadrons is produced. The gluon polarization
$\Delta G/G$ is extracted from this asymmetry using 
the event generator PYTHIA 6.2~\cite{Sjostrand:2001yu} and leading-order
expressions for the analyzing powers of the PGF and of the background
processes.
Possible spin effects in the fragmentation are neglected.

\section{Experimental set-up}

The experiment \cite{Mallot:2004gk} is located at the M2 beam line of the 
CERN SPS, which provides a 160 GeV $\mu^+$ beam at a rate of $2 \times
10^8$ muons per spill of 4.8~s with a cycle time of 16.8~s. The muons
are produced in the decay of pions and kaons, and the beam has a natural
polarization of $\langle P_{\rm b} \rangle = -0.76$, with a relative
accuracy of $5\%$~\cite{Ageev:2005g1}. The incident muon momentum is
measured upstream of the experimental area in a beam spectrometer, while
its direction and position at the entrance of the target are determined
in a telescope of scintillating fiber hodoscopes and silicon microstrip
detectors.

The polarized target system \cite{Ball:2003vb} consists of an upstream
cell (u) and a downstream cell (d), each 60 cm long and 3 cm in
diameter, separated by 10 cm.  The cells are located on the axis of a
superconducting solenoid magnet providing a field of 2.5 T along the
beam direction, and are filled with $^6$LiD. This material is used as a
deuteron target and was selected for its high dilution factor $f$ of
about 40\%, which accounts for the fact that only a fraction of the
target nucleons are polarizable. Typical polarization values of 50\% are
obtained by dynamic nuclear polarization, and measured with a relative
accuracy of $5$\%. 
The two cells are polarized in opposite directions by using different
microwave frequencies so that data with both spin orientations
are recorded simultaneously. The muon flux then cancels out in the
counting rate asymmetry. However, the acceptance of the
spectrometer is not identical for the two cells, which gives rise to an
acceptance asymmetry. To account for this, a rotation of the magnetic
field is performed in order to reverse the orientation of the spins in
each cell. The acceptance asymmetry then disappears in the sum between
the counting rate asymmetries before and after rotation
(for details, see Eq.~2).  A perfect cancellation requires the ratio $L_{\rm u} a_{\rm u}/L_{\rm d} a_{\rm d}$
to be the same before and after rotation, where $L_{\rm u},L_{\rm d}$
are the luminosities and
$a_{\rm u}, a_{\rm d}$ the acceptances for the upstream and downstream
target cells. False asymmetries due to the variations of this ratio with
time are
minimized by performing the rotation frequently, i.e.~every 8 hours.
However, because of the change in the orientation of the target field,
the set-up is slightly different before and after rotation, which
affects the $L_{\rm u} a_{\rm u}/L_{\rm d} a_{\rm d}$ ratio. To
cancel this effect the orientation of the spins for a given field
orientation is reversed by repolarization a few times during the running period.

The COMPASS spectrometer has a large angle and a small angle
spectrometer built around two dipole magnets, in order to allow the
reconstruction of the scattered muon and of the produced hadrons in
broad momentum and angular ranges.
Different types of tracking detectors are used to deal with the rapid
variation of the particle flux density with the distance from the beam.
Tracking in the beam region
is performed by scintillating fibers. Up to $20~$cm from the beam we use
Micromegas and GEMs. Further away, tracking is carried out in multiwire
proportional 
chambers and drift chambers. Large-area trackers, based on straw
detectors and large drift chambers
extend the tracking over a surface of up to several square meters. Muons
are identified by dedicated trackers 
placed downstream of hadron absorbers. Hadron/muon separation is
strengthened by two large iron--scintillator sampling calorimeters,
installed upstream of the hadron absorbers and shielded to avoid
electromagnetic contamination. The particle identification provided by
the ring imaging Cherenkov detector is not used in the present analysis.

The trigger system~\cite{trigger} provides efficient tagging down to
$Q^2 = 0.002$~$({\rm GeV}/c)^2$, by detecting the scattered muon in a
set of hodoscopes placed behind the two dipole magnets. 
A large enough energy deposit in the hadronic calorimeters is required
in order to suppress unwanted triggers 
generated by halo muons, elastic muon--electron scattering events, and radiative events.
\section{Asymmetry measurement}

The present analysis deals with data collected in 2002 and 2003. The
selected events are required to contain at least two charged hadrons
associated to the primary vertex, in addition to the incident and
scattered muons. We consider events with $0.35 < y < 0.9$, where $y$ is
the fraction of energy lost by the incident muon. The lower $y$ cut
removes events with a low sensitivity to the gluon polarization, while
the upper one rejects events which could be affected by large radiative
effects. Since PYTHIA provides a reliable model for interactions of
virtual photons with nucleons at low virtuality~\cite{Friberg:2000ra},
we select events with $Q^2<1$~$({\rm GeV}/c)^2$, which corresponds to
about 90\% of the total data set. The DIS sample, $Q^2>1$~$({\rm
GeV}/c)^2$, is being analyzed separately using LEPTO which is better
adapted to this domain.

Furthermore, cuts are applied on the two hadrons with highest transverse
momentum. The muon contamination of the hadron sample is eliminated by
requiring the energy deposit in the calorimeters to be large enough with
respect to the reconstructed momentum, $E/p>0.3$. In addition, hadron
candidates detected also downstream of the hadron absorbers are
discarded. The invariant mass of the two-hadron system is required to be
larger than 1.5~${\rm GeV}/c^2$, and $x_{\rm F}$ to be larger than 0.1,
where $x_{\rm F}=2 p_{\rm L}^*/W$. Here, $p_{\rm L}^*$ is the
longitudinal momentum of the hadron in the photon--nucleon center of
mass frame and $W$ is the invariant mass of the hadronic final state.
Finally, the fraction of PGF events in the sample is enhanced by
requiring the transverse momentum of the two hadrons to be large:
$p_{\rm T}^{\rm h1}>0.7$~GeV$/c$, $p_{\rm T}^{\rm h2}>0.7$~GeV$/c$ and
$(p_{\rm T}^{\rm h1})^2 + (p_{\rm T}^{\rm h2})^2>2.5$~$({\rm GeV}/c)^2$,
as in the SMC high-$p_{\rm T}$ analysis~\cite{Adeva:2004dh}. In total,
around 250\,000 events remain after these cuts, defining the
high-$p_{\rm T}$ sample.

The asymmetry $A_\parallel$ can be obtained from the number of events in
the upstream and downstream cells, before and after field rotation:
\begin{equation}
A_\parallel = \frac{1}{2 | P_{\rm b} P_{\rm t} f | } \left(
\frac{N_{\rm u}^{\uparrow \Downarrow} - N_{\rm d}^{\uparrow \Uparrow}}{N_{\rm u}^{\uparrow \Downarrow} + N_{\rm d}^{\uparrow \Uparrow}}
+  
\frac{N_{\rm d}^{\uparrow \Downarrow} - N_{\rm u}^{\uparrow \Uparrow}}{N_{\rm d}^{\uparrow \Downarrow} + N_{\rm u}^{\uparrow \Uparrow}}
\right).
\label{eq:Apar}
\end{equation}
The two terms in this expression correspond to opposite orientations of
the target magnetic field, with for example $N_{\rm u}^{\uparrow
\Downarrow}$ the number of events in the upstream cell when the cell
polarization is anti-parallel to the beam polarization.

The statistical error on the asymmetry is minimized by weighting each
event with its overall sensitivity to the gluon
polarization~\cite{Adams:1997smc}. The event weight is taken to be $w$ =
$fDP_{\rm b}$, where $D$ is a kinematic factor which approximates the
amount of polarization transferred from the incident muon to the virtual
photon:
\begin{equation}
D = \frac{y ( 2 - y -\frac{2 m^2 y^2}{Q^2 (1-x  y)})}{( 1+ (1-y)^2-\frac{2 m^2 y^2}{Q^2})\sqrt{1-\frac{4 m^2 (1-x) x y^2}{Q^2 (1-x y)^2}}}.
\label{eq:depolfact}
\end{equation}
Here, all terms containing the muon mass $m$ were taken into account
since the sample of events is at low $Q^2$. The factor $D$ is
proportional to the analyzing power of PGF apart from a weak dependence
on the event kinematics, and was therefore used in the weight instead of
$\hat a_{\rm LL}^{\rm PGF}$ which is unknown on an event-by-event basis.
The average  value of $D$ is around 0.6. In the weighting method, the
expression for the asymmetry becomes
\begin{equation}
\left \langle \frac{A_\parallel}{D} \right \rangle= 
\frac{1}{2  |P_{\rm t}| } 
\left(
\frac{\Sigma w_{\rm u}^{\uparrow \Downarrow} - \Sigma w_{\rm d}^{\uparrow \Uparrow}}
{\Sigma (w_{\rm u}^{\uparrow \Downarrow})^2 + \Sigma (w_{\rm d}^{\uparrow \Uparrow})^2}
+
\frac{\Sigma w_{\rm d}^{\uparrow \Downarrow} - \Sigma w_{\rm u}^{\uparrow \Uparrow}}
{\Sigma (w_{\rm d}^{\uparrow \Downarrow})^2 + \Sigma (w_{\rm u}^{\uparrow \Uparrow})^2}
\right).
\label{eq:Aparweight}
\end{equation}
With the  high-$p_{\rm T}$ sample defined above, we obtain
\begin{equation}
\left \langle \frac{A_\parallel}{D} \right \rangle= 0.002 \pm 0.019 {\rm (stat)} \pm 0.003 {\rm (syst)}.
\label{eq:measaparodall}
\end{equation}
The systematic error accounts for the false asymmetries, which were
estimated using a sample of low $p_{\rm T}$ events with much larger
statistics. Other sources of systematic errors, including the error on
the beam and target polarizations, are proportional to the (small)
measured asymmetry, and have been neglected.
\section{Gluon polarization}

As stated before, the determination of the gluon polarization from the
high-$p_{\rm T}$ asymmetry involves a Monte Carlo simulation. 
The generated events are propagated through a GEANT~\cite{geant} model
of the COMPASS spectrometer, and reconstructed using the same program as
for real data. Finally, the same cuts as for real data are applied to
obtain the Monte Carlo sample of high-$p_{\rm T}$ events. 

We use PYTHIA to generate two different kinds of processes. In direct
processes, for example the PGF, the virtual photon takes part directly
in the hard partonic interaction. In resolved-photon processes, it
fluctuates into a hadronic state from which a parton is extracted (the
partonic structure of the virtual photon is resolved). This parton then
interacts with a parton from the nucleon. At $Q^2<1$~$({\rm GeV}/c)^2$,
the resolved-photon processes constitute about half of the high-$p_{\rm
T}$ sample. For $Q^2>1$~$({\rm GeV}/c)^2$, their contribution drops to
about 10\%, and it becomes negligible for $Q^2 > 2$~$({\rm GeV}/c)^2$.
To compute the asymmetry of a given process, it is mandatory to find a
hard scale $\mu^2$ allowing the factorization of the cross-sections into
a hard partonic cross-section calculable perturbatively, and a soft
parton distribution function which needs to be measured. In
Eq.~(\ref{eq:aparod_decomppgf}) for instance, the asymmetry of the PGF
factorizes into a hard asymmetry (the analyzing power) and a soft
asymmetry (the gluon polarization). In our case, $Q^2$ is too small to
be used as a scale. However, the scale provided by
PYTHIA\footnote{Parameter MSTP(32).} is very close to the $p_{\rm T}^2$
of one of the partons produced in the hard reaction. Since the $p_{\rm
T}$ cut applied to the two highest-$p_{\rm T}$ hadrons implies, for most
of the events, large transverse momentum partons in the final state of
the hard reaction, this quantity turns out to be large enough. Events
for which no hard scale can be found are classified in PYTHIA as
``low-$p_{\rm T}$ processes.''

After varying many parameters of PYTHIA, the best agreement with our
data was obtained by modifying only 
the width of the intrinsic transverse momentum distribution of partons
within the resolved virtual photon\footnote{Parameter PARP(99) in
Ref.~\cite{Sjostrand:2001yu}. }, which was decreased from 1~GeV$/c$ to
0.5~GeV$/c$. 

\begin{figure}[h!]
\centering
\includegraphics[width=0.49\textwidth]{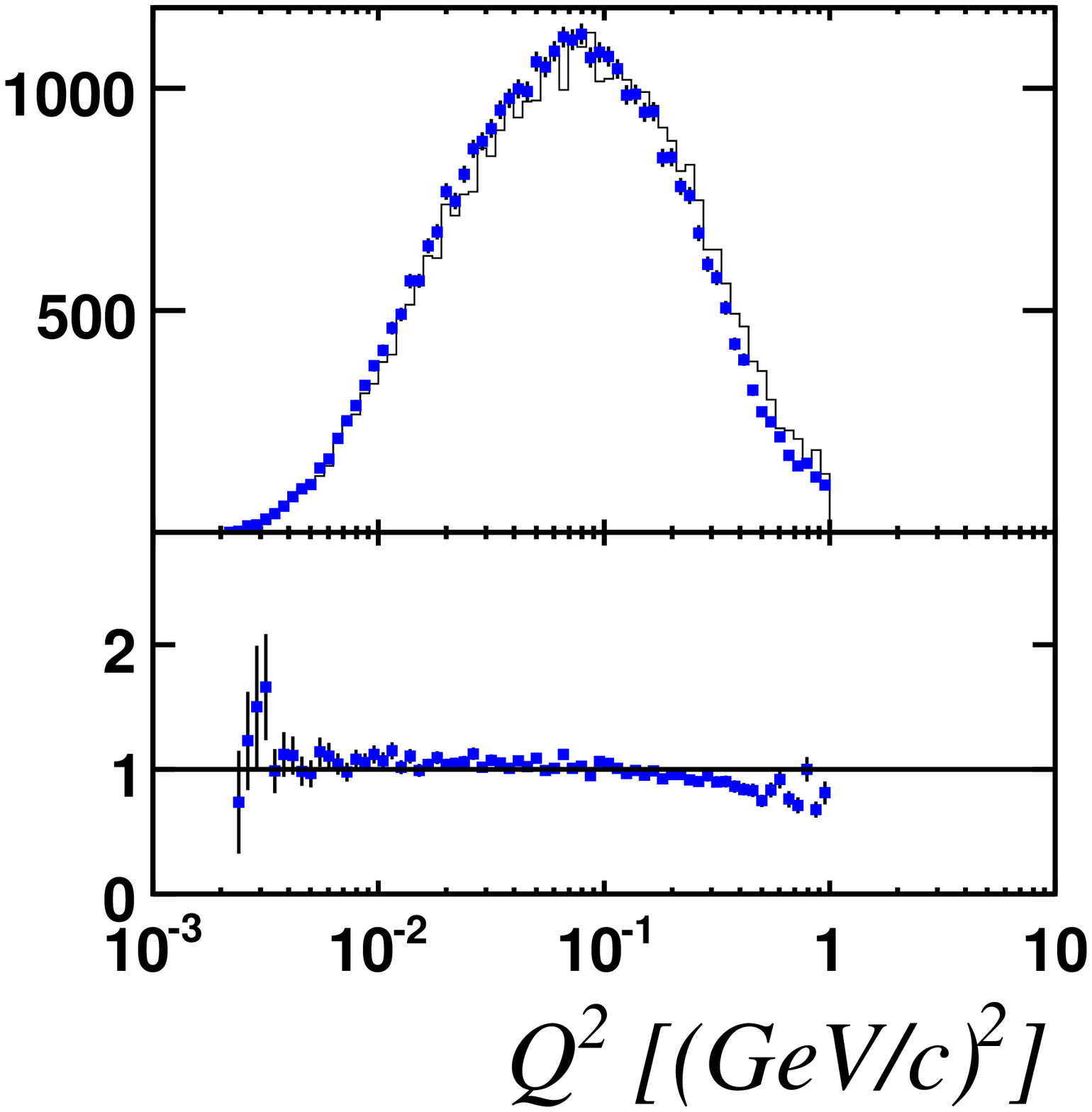}
\includegraphics[width=0.49\textwidth]{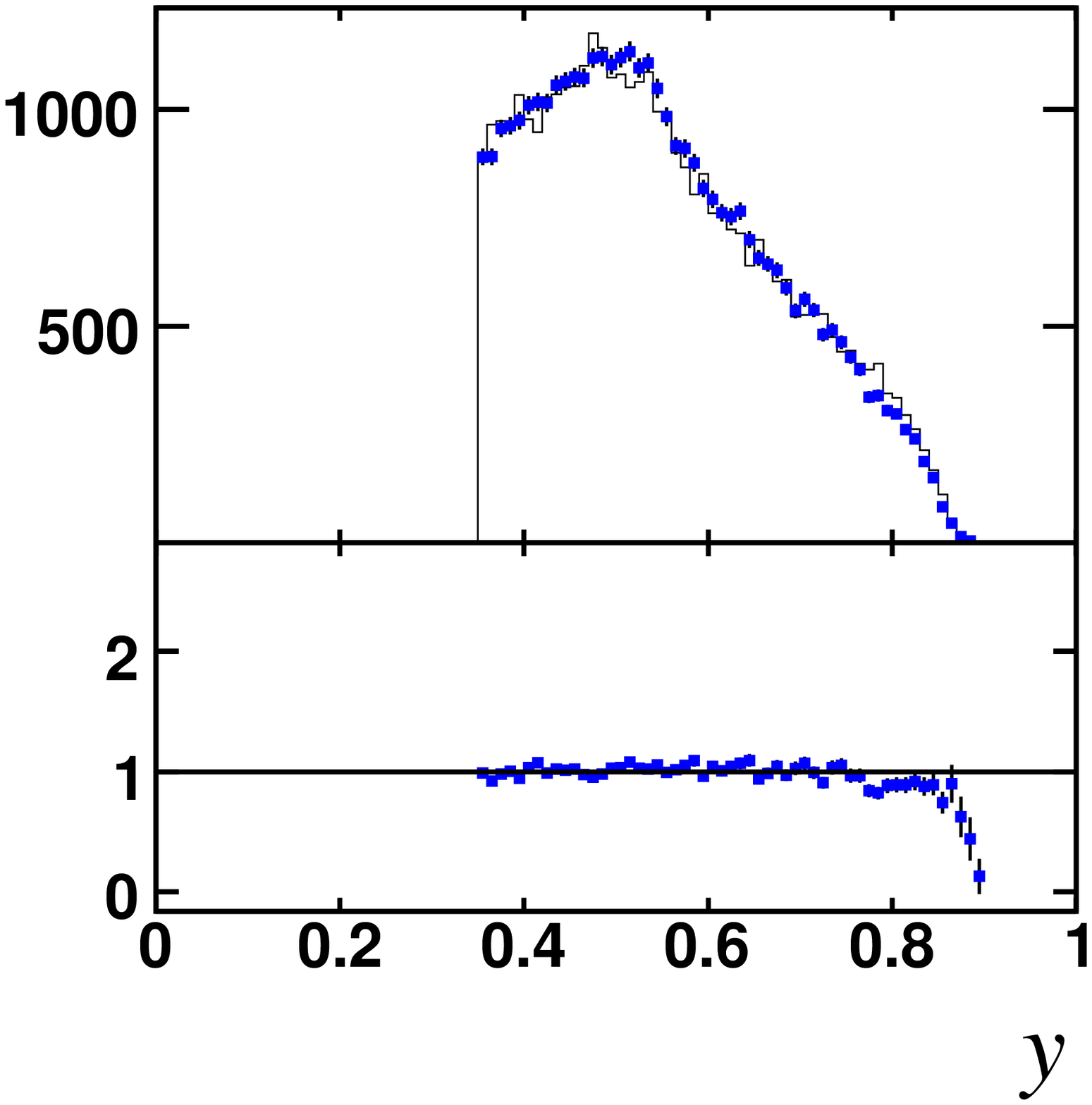}\\
\includegraphics[width=0.49\textwidth]{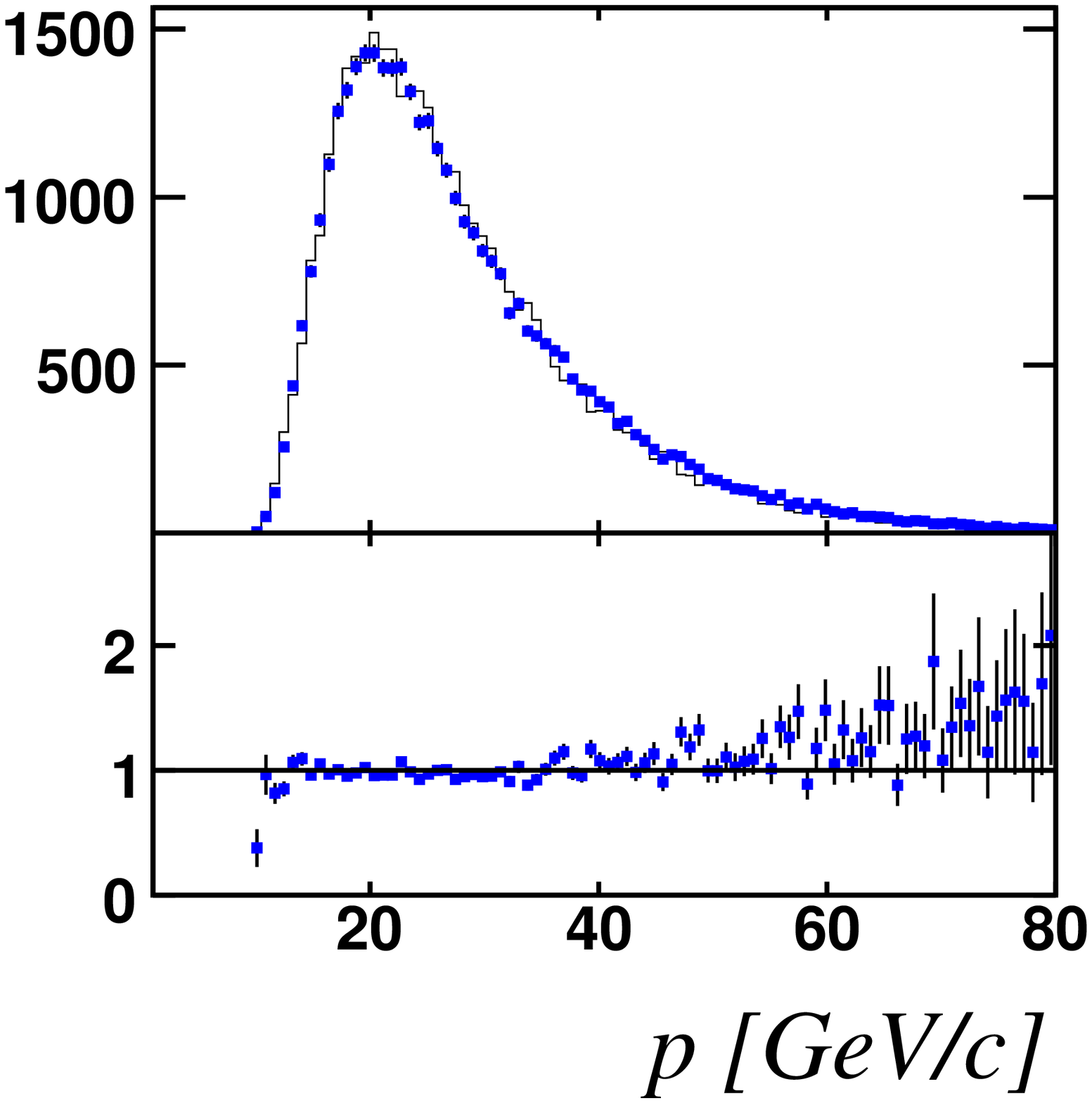}
\includegraphics[width=0.49\textwidth]{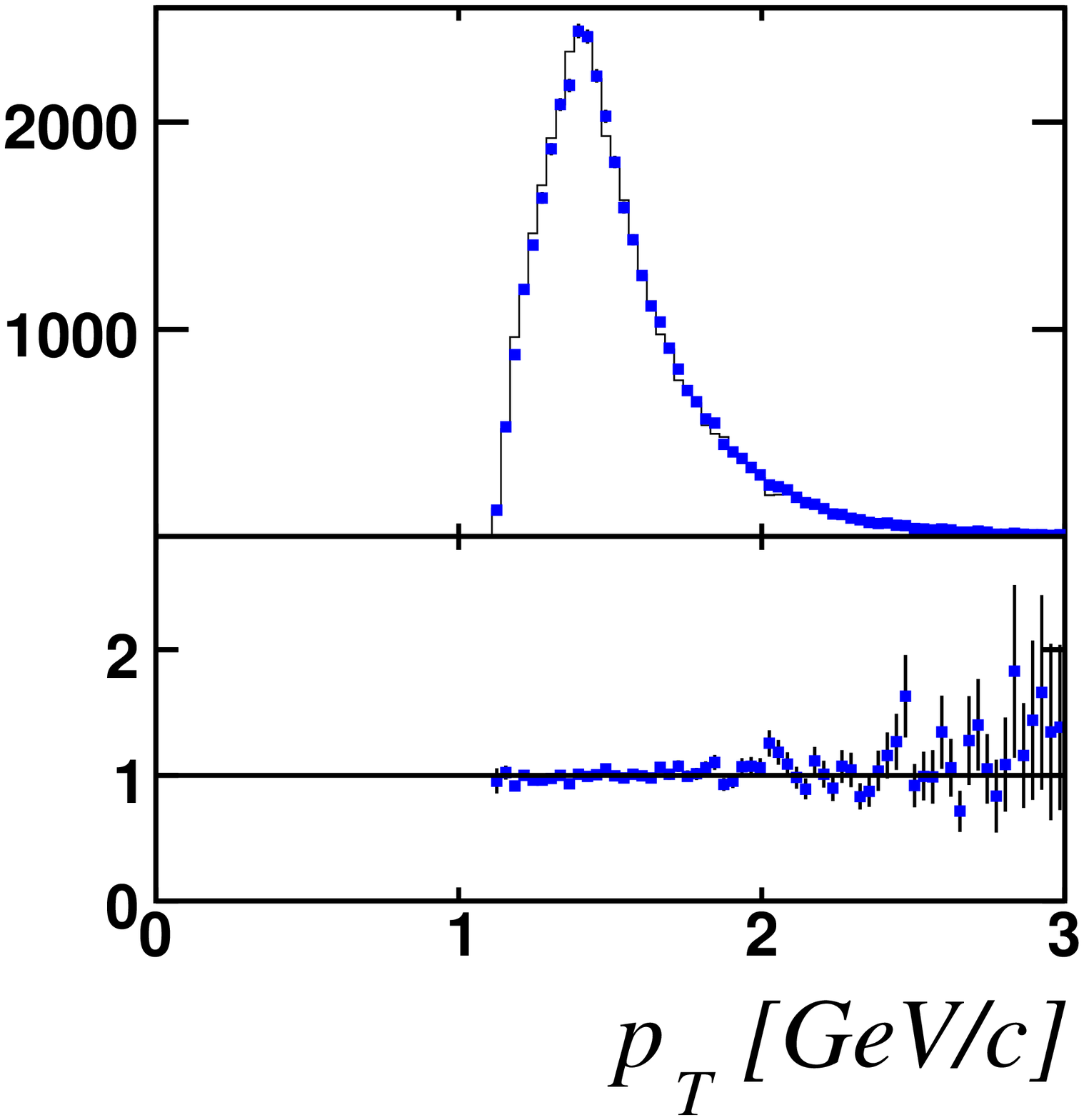}
\caption{Comparison between data and Monte Carlo for $Q^2$, $y$, and for
the total (transverse) momentum $p$ ($p_{\rm T}$) of the hadron with
highest $p_{\rm T}$. The upper part of each plot shows the  real data
(points) and simulation (line), normalized to the same number of events.
The lower part shows the corresponding data/Monte Carlo ratio. }
\label{fig:mcrealcomp}
\end{figure}

The lower $p_{\rm T}$ cut-off\footnote{Parameter CKIN(5).} is set by
default to 1~GeV$/c$ to prevent the cross-section for $2\rightarrow 2$
processes such as PGF from diverging when the partonic transverse
momentum vanishes. However, this does not occur in our high-$p_{\rm T}$
sample, as the transverse-momentum distribution of the outgoing partons
starts just below 1~GeV$/c$. To avoid cutting into this distribution, we
have decreased the lower $p_{\rm T}$ cut-off to 0.9~GeV$/c$ and did not
observe any effect on the agreement with the data.
The simulated and real data samples of high-$p_{\rm T}$ events are
compared in Fig.~\ref{fig:mcrealcomp} for $Q^2$, $y$, and for the total
and transverse momenta of the hadron with highest $p_{\rm T}$. An
equally good agreement is obtained for the second hadron.

\begin{figure}
\centering
\includegraphics[width=\textwidth]{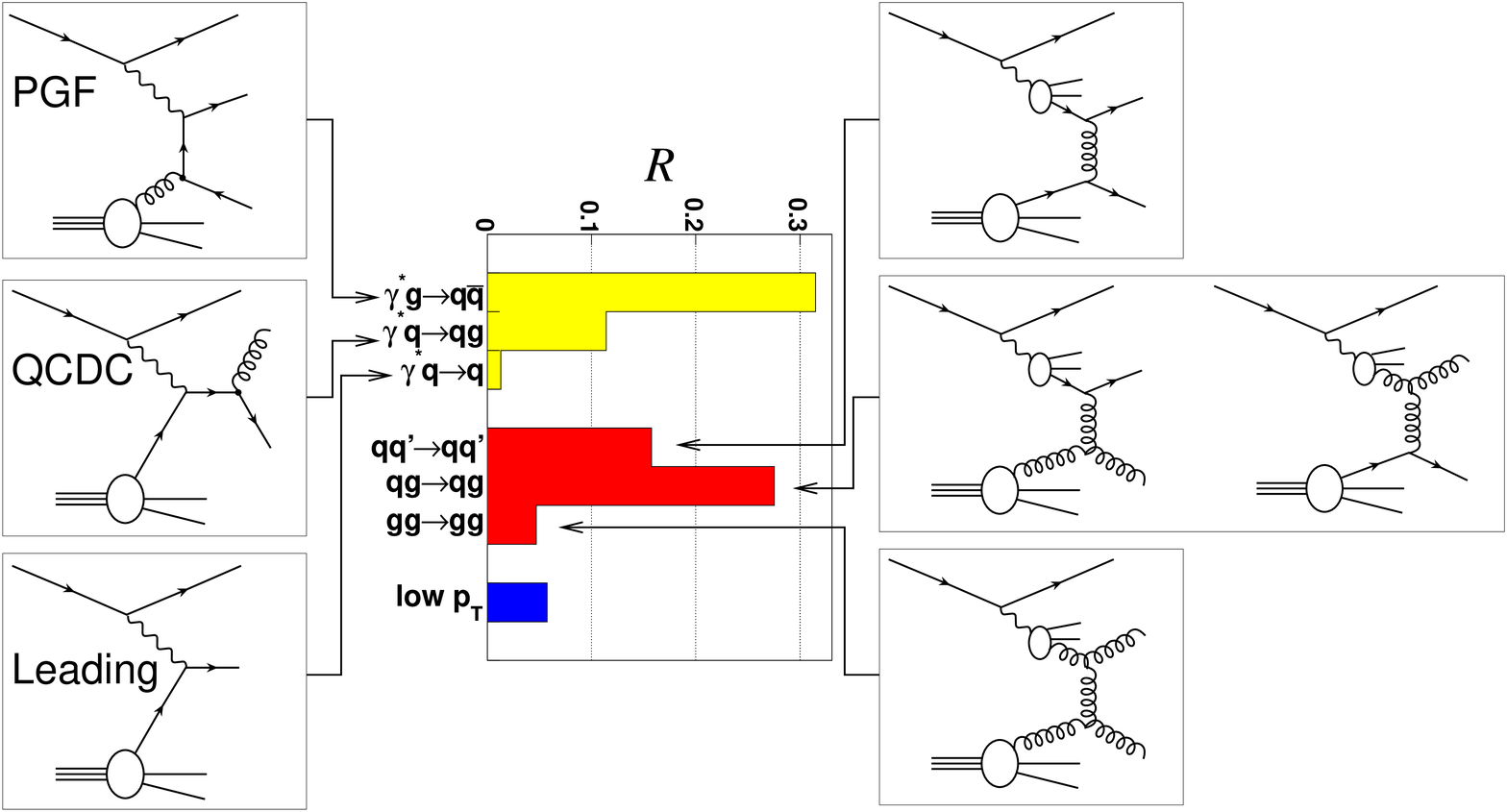}
\caption{Relative contributions $R$ of the dominant PYTHIA processes to
the Monte Carlo sample of high-$p_{\rm T}$ events. Left: direct
processes; right: resolved-photon processes.}
\label{fig:rsubproc}
\end{figure}

Various processes contribute to the Monte Carlo sample of high-$p_{\rm
T}$ events, as shown in Fig.~\ref{fig:rsubproc}.  The direct processes
are the PGF, the QCD Compton (QCDC, $\gamma^* q \rightarrow q g $), and
the leading process ($\gamma^* q \rightarrow q$). For the
resolved-photon processes, a parton $f$ from the nucleon interacts with
a parton $f^\gamma$ from the virtual photon, where $f$ and $f^\gamma$
can be a quark or a gluon. We have neglected the resolved-photon
processes $q \bar q \rightarrow q' \bar q'$, $q \bar q \rightarrow gg$
and $gg \rightarrow q \bar q$, which altogether represent only $0.6\%$
of the sample. The low-$p_{\rm T}$ processes contain all events for
which no hard scale can be found. Each of these processes contributes to
the cross-section helicity asymmetry, provided that a transverse photon
is exchanged. The asymmetry can then be approximately expressed as
\begin{align}
  \nonumber \left \langle \frac{A_\parallel}{D} \right \rangle &= R_{\rm PGF} 
            \left\langle \frac{\hat a_{\rm LL}^{\rm PGF}}{D} \right\rangle
\frac{\Delta G}{G}
  + R_{\rm QCDC} \left \langle \frac{\hat a_{\rm LL}^{\rm QCDC}}{D} A_1 \right \rangle \\
  \nonumber &+ \sum_{f,f^\gamma=u,d,s,\bar u, \bar d, \bar s, G}
R_{ff^\gamma} \left\langle \hat a_{\rm LL}^{ff^\gamma}  \frac{\Delta
f}{f} \frac{\Delta f^\gamma}{f^\gamma} \right\rangle\\
  &+ R_{\rm leading} \times A_{\rm leading} + R_{\rm low-p_T} \times A_{\rm low-p_T}.
\label{eq:allcontribs}
\end{align}
Here, $R_{\rm QCDC}$ is the fraction of QCD Compton events, and
$R_{ff^\gamma}$ is the fraction of events in the whole high-$p_{\rm T}$
sample for which a parton $f$ from the nucleon interacts with a parton
$f^\gamma$ from a resolved photon. Recalling that we use a deuteron
target, $A_1$ is the inclusive virtual-photon--deuteron asymmetry and
$\Delta f/f$ ($\Delta f^\gamma/f^\gamma$) is the polarization of quarks
or gluons in the deuteron (photon). The contributions of the leading and
low-$p_{\rm T}$ processes cannot be calculated in the same way, since
there is no hard scale allowing the factorization of their asymmetries
$A_{\rm leading}$ and $A_{\rm low-p_T}$ (low transverse momentum, and
$Q^2<1$~$({\rm GeV}/c)^2$ events). However, the asymmetry for this kind
of events is small as indicated by previous measurements of $A_1$ at low
$Q^2$~\cite{Adeva:1999pa}. Moreover, the leading and low-$p_{\rm T}$
processes together account for only 7\% of the high-$p_{\rm T}$ sample.
For these two reasons, we neglected their contributions.
  
The fraction of photon--gluon fusion events in the sample is of the order of 30\%,
see Fig.~\ref{fig:rsubproc}. The analyzing power $\hat a_{\rm LL}^{\rm
PGF}$ is calculated using the leading-order expressions for the
polarized and unpolarized partonic cross-sections and the parton
kinematics for each PGF event in the high-$p_{\rm T}$ Monte Carlo
sample. In average, we obtain $\left\langle \hat a_{\rm LL}^{\rm PGF}/D
\right\rangle = -0.93$, so that   
the contribution of PGF to the high-$p_{\rm T}$ asymmetry is $-0.29 \times \Delta G / G$.

The contribution of QCD Compton events to the high-$p_{\rm T}$ asymmetry
is evaluated from a parametrization of the virtual-photon--deuteron
asymmetry $A_1$ based on a fit to the world
data~\cite{Adeva:1998vv,Abe:1998wq}. This asymmetry is calculated for
each event at the momentum fraction $x_{\rm q}$ of the quark, known in
the simulation. The estimated contribution of the QCD Compton scattering
to the high-$p_{\rm T}$ asymmetry is 0.006.

The parton from a resolved photon interacts either with a quark or a
gluon from the nucleon. In the latter case, the process is sensitive to
the gluon polarization $\Delta G/G$. The analyzing powers $\hat a_{\rm
LL}^{ff^\gamma}$ are calculated in pQCD at leading
order~\cite{Bourrely:1987gp}. The polarizations of the $u$, $d$ and $s$
quarks in the  deuteron $\Delta f/f$ are calculated using 
the polarized parton distribution functions from Ref.~\cite{Gluck:2000dy} (GRSV2000) and
the unpolarized parton distribution functions from
Ref.~\cite{Gluck:1998xa} (GRV98, also used as an input for PYTHIA), all
at leading order. The polarizations of quarks and gluons in the virtual
photon $\Delta f^\gamma/f^{\gamma}$ are unknown because  the polarized
PDFs of the virtual photon  have not yet been measured. Nevertheless,
theoretical considerations provide a minimum and a maximum value for
each $\Delta f^\gamma$,
in the so-called  minimal and  maximal scenarios~\cite{Gluck:2001rn}. As
the analyzing powers are positive for all considered channels, the two
scenarios correspond to extreme values for the contribution of the
resolved-photon processes to the high-$p_{\rm T}$ asymmetry, $0 + 0.012
\times \Delta G/G$ and $0.002 + 0.078 \times \Delta G/G$, respectively.
Here, the term proportional to $\Delta G/G$ comes from the processes
involving a gluon from the nucleon.

Our analysis is restricted to leading order, and the systematic error
has to take into account next-to-leading-order effects. Their order of
magnitude is estimated by repeating the analysis several times with
modified Monte Carlo parameters: 
the renormalization and factorization scales were multiplied and divided
by two, and the parton shower mechanism was deactivated. The systematic
error is obtained from the difference in the corresponding values for
$\Delta G/G$, 0.004 (0.011) in the minimal (maximal) scenario. 

Another source of systematic error is the tuning of the PYTHIA
parameters. Since our event selection relies on a cut in transverse
momentum, the relevant parameters are those which determine the amount
of transverse momentum acquired by the outgoing hadrons in the soft
parts of the reaction: the intrinsic transverse momentum of the partons
in the nucleon and in the resolved photon, and the parameters describing
the hadronization. These parameters were scanned independently over a
range in which the  agreement between the simulation and real data
remains reasonable. This results in several values for $\Delta G/G$, all
based on the same measured high-$p_{\rm T}$ asymmetry of
Eq.~(\ref{eq:measaparodall}). 
The value of $\Delta G/G$ appears to depend predominantly on the width
of the intrinsic-transverse-momentum distribution for the partons in the
photon. 
Varying this parameter between 0.1~GeV$/c$ and 1~GeV$/c$ leads to a 30\%
variation of the fraction of photon--gluon fusion $R_{\rm PGF}$.
Note that the resulting systematic errors are proportional
to the high-$p_{\rm T}$ asymmetry, which implies that a statistical
fluctuation of the measured high-$p_{\rm T}$ asymmetry modifies the
systematic errors. This was taken into account by performing the
systematic study for $A_\parallel/D + \sigma_{\rm stat} (A_\parallel/D)$
and $A_\parallel/D - \sigma_{\rm stat} (A_\parallel/D)$ as well, quoting
the largest value for the systematic error.
Finally, the systematic error on $\Delta G/G$ is  0.018 and 0.052 in the
minimal and maximal scenarios, respectively. 
\section{Result and conclusion}

The values for the gluon polarization in the minimal and maximal scenarios are 
\begin{align}
\left( \frac{\Delta G}{G}\right)_{\rm min} &= 0.016 \pm 0.068 {\rm (stat)} \pm 0.011 {\rm (exp.\ syst)} \pm 0.018 {\rm (MC.\ syst)},
\label{eq:dgogmin20023}\\
\left( \frac{\Delta G}{G}\right)_{\rm max} &= 0.031 \pm 0.089 {\rm (stat)} \pm 0.014 {\rm (exp.\ syst)} \pm 0.052 {\rm (MC.\ syst)}.
\label{eq:dgogmax20023}
\end{align}
This leads to the central value
\begin{equation}
\frac{\Delta G}{G} = 0.024 \pm 0.089 {\rm (stat)} \pm 0.057 {\rm (syst)}, 
\label{eq:dgog20023quadsys}
\end{equation}
where the difference between the two scenarios has been included in the
systematics, and where all systematics have been added quadratically.
Let us recall that the systematic error covers an uncertainty on $R_{\rm
PGF}$ of up to 30\%.
Gluons are probed at an average scale $\mu^2$ and an average momentum fraction of the gluons $x_{\rm g}$, which are both obtained from the simulation where the parton kinematics is known. 
For the scale, we obtain in average $\mu^2 \simeq 3$~$({\rm GeV}/c)^2$.
The distribution of $x_{\rm g}$ is asymmetric, with a different r.m.s.
width on the left and on the right, $x_{\rm g}=0.095^{+0.08} _{-0.04}$.
For these two quantities, the average was obtained by weighting each
event by its sensitivity to the gluon polarization,
c.f.~Eq.~(\ref{eq:allcontribs}).

\begin{figure}
\centering
\includegraphics[width=0.8\textwidth]{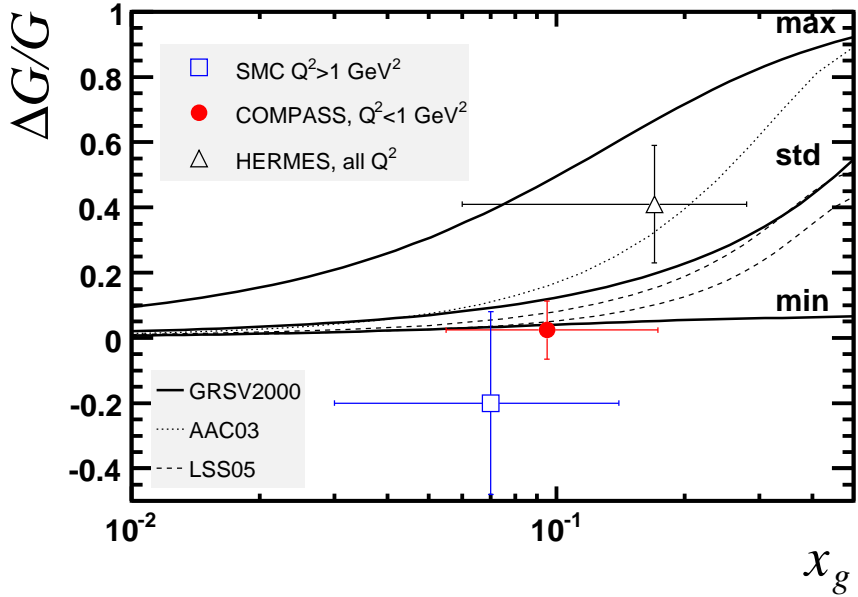}
\caption{Comparison of the $\Delta G/G$ measurements from COMPASS
(present work), SMC~\cite{Adeva:2004dh}, and
HERMES~\cite{Airapetian:1999ib}. The horizontal bar on each point
represents the range in $x_{\rm g}$. The curves show various
parametrizations from NLO fits in the $\overline{\rm MS}$ scheme at
$\mu^2 = 3$ (GeV/c)$^2$: GRSV2000~\cite{Gluck:2000dy} (3 curves, please
see text for details), AAC03~\cite{Hirai:2003pm}, and  LSS05 sets 1 and
2~\cite{Leader:2005kw}. }
\label{fig:dggcomp}
\end{figure}

Our value for the gluon polarization is compared with previous direct
measurements from the SMC~\cite{Adeva:2004dh} and
HERMES~\cite{Airapetian:1999ib} experiments in Fig.~\ref{fig:dggcomp}.
The SMC measurement uses high-$p_{\rm T}$ events at $Q^2 > 1$~$({\rm
GeV}/c)^2$, where the contribution of the resolved-photon processes is
small. 
The HERMES result was derived from data mostly at low $Q^2$, but the
contribution of the resolved-photon processes to the asymmetry was
neglected. Note that these experimental determinations are based on a
leading-order analysis.
                                                                                                              
Figure~\ref{fig:dggcomp} also shows three distributions of $\Delta G/G$
as a function of $x_{\rm g}$
from Ref.~\cite{Gluck:2000dy} (GRSV2000),
resulting from QCD fits to the world $g_1$ data at NLO.
They  correspond to three hypotheses on the gluon
polarization at $\mu^2 = 0.40$~$({\rm GeV}/c)^2$: maximal polarization
(max), best fit to the data (std), and zero polarization (min). The
distributions are then evolved radiatively to $\mu^2 = 3$~$({\rm
GeV}/c)^2$ where their first moments are 2.5, 0.6 and 0.2, respectively.
Our result clearly favors parametrizations with a low gluon
polarization.

More recent NLO distributions of $\Delta G/G$ 
from Ref.~\cite{Hirai:2003pm} (AAC03) and Ref.~\cite{Leader:2005kw}
(LSS05, sets 1 and 2) are displayed as well.
Although these curves
strongly differ in shape, the values at $x_{\rm g} = 0.095$ are quite
close and all within $1.5~\sigma$ above our measured value. 
The first moment $\Delta G$ at the scale $\mu^2 = 3$~$({\rm GeV}/c)^2$
is equal to $0.8$ for AAC03, and to 0.26 (0.39) for LSS05 set 1 (set 2).

When the singlet axial matrix element $a_0$ was found to be much smaller than
the contribution to the nucleon spin expected in the naive quark-parton
model, it was suggested that the difference could be accounted for by a
large contribution of the gluon
spin~\cite{Efremov:1988zh,Altarelli:1988nr,Carlitz:1988ab}. Indeed, in
the so-called
AB~\cite{Ball:1995td} or JET~\cite{Leader:1998nh} renormalization schemes, the contribution of the
quark spins to the nucleon spin becomes $\Delta \Sigma = a_0 + N_f (\alpha_s/2 \pi) \Delta G$
where $N_f$ is the number of active flavors. At $Q^2 = 3$~$({\rm GeV}/c)^2$, a value
of $\Delta G$ of about 3 would be required to obtain the expected $\Delta \Sigma$
of the order of 0.6.
The small value of $\Delta G/G$ at $x_{\rm g} = 0.095$ from our measurement cannot by itself
rule out the possibility of the first moment $\Delta G$ being as large as 3,
since the shape of $\Delta G(x_{\rm g})$ is poorly known. However, the
fact that our point is lower than fitted parameterizations, leading to
values of
$\Delta G$ around or below unity, makes the hypothesis of a large $\Delta G$ unlikely.

In summary, we have measured the gluon polarization at $x_{\rm g}
=0.095$ and $\mu^2 \simeq 3$~$({\rm GeV}/c)^2$ and found a result
compatible with zero, with a statistical error and a systematic error
smaller than 0.1. The gluon polarization was extracted from the
longitudinal spin asymmetry obtained for low-$Q^2$ events in which a
pair of high-$p_{\rm T}$ hadrons is produced. 
The present analysis, for the first time, takes into account the
contribution from the polarized structure of the virtual photon.\\

{\bf Acknowledgments}

We gratefully acknowledge the support of the CERN management and staff
and the skill and effort of the technicians of our collaborating
institutes. Special thanks are due to V.~Anosov and V.~Pesaro for their
technical support during the installation and the running of this
experiment. We also thank B. Pire, M. Stratmann, and W. Vogelsang for
useful discussions. 

\end{document}